\documentclass{PoS}
\title{Discovery of VHE gamma-rays from the radio galaxy PKS\,0625-354 with H.E.S.S.}

\ShortTitle{PKS\,0625-354 with H.E.S.S.}

\author{\speaker{Michal Dyrda}$^a$, Alicja Wierzcholska$^b$$^,$$^a$\footnote{Mobility Plus Fellow}, Olivier Hervet$^c$, Rafal Moderski$^d$, Mateusz Janiak$^d$, Micha{\l} Ostrowski$^e$ and {\L}ukasz Stawarz$^e$ for the H.E.S.S. Collaboration\\
        \llap{$^a$} Institute of Nuclear Physics, PAS, ul. Radzikowskiego 152, 31-342 Krak\'ow, Poland\\
	\llap{$^b$} Landessternwarte, Universit\"at Heidelberg, K\"onigstuhl 12, D 69117 Heidelberg, Germany\\
	\llap{$^c$} LUTH, Observatoire de Paris, CNRS, Universit Paris Diderot, 5 Place Jules Janssen, 92190 Meudon, France\\
	\llap{$^d$} Nicolaus Copernicus Astronomical Center, ul. Bartycka 18, Warsaw, Poland\\
	\llap{$^e$} Astronomical Observatory of the Jagiellonian University, ul. Orla 171, 30-244 Krak\'ow, Poland
        E-mail: \email{Michal.Dyrda@ifj.edu.pl}}

\abstract{Most of the extragalactic objects detected so far in the very high energy (VHE) regime are blazars, but the discovered nearby radio galaxies: M87, Cen A and NGC 1275 of type FRI seem to constitute a new class of VHE emitters. The radio galaxy PKS\,0625-354 was observed and detected ($\sim$6$\sigma$) with the H.E.S.S. phase I telescopes in 2012, above an energy threshold of 250 GeV. The time-averaged VHE energy spectrum is well characterized by a power law model. The broad-band light curve, including the available multiwavelength data, as well as the VHE data gathered with H.E.S.S. will be presented.}

\FullConference{The 34th International Cosmic Ray Conference,\\
		30 July- 6 August, 2015\\
		The Hague, The Netherlands}

\begin{document}

\section{Introduction}
Blazars are the most numerous class of extragalactic objects discovered in the very high energy (VHE; E $>$ 100 GeV). These are active galactic nuclei with jets pointing close to the line of sight.  However, there is a growing evidence that blazars are not the only extragalactic objects capable of VHE emission. With the detection of M 87, Cen A, IC 310 and Per A nearby radio galaxies of type FRI seem to constitute a new class of VHE sources \cite{M87,CenA,Hildebrand}.

Radio galaxies (RGs) are active galaxies with their relativistic jets oriented at intermediate to larger viewing angles with respect to the line of sigh\cite{UrryPadovani}. As a result of larger inclinations, the observed non-thermal emission produced within the innermost parts of the jets is not amplified by relativistic beaming and hence different emission components, typically not present in observed blazar spectra, may become prominent. 

Increasing the number of VHE ''$\gamma$-ray loud'' RGs is important for several reasons. First, modelling of such sources provides an independent check of blazar models which are being developed, since RGs are considered as blazars observed at larger viewing angles \cite{Abdo2010,Fukazawa}. Second, $\gamma$-ray observations of RGs may reveal some ''exotic'' or at least non-standard processes possibly related to the production of high energy photons and particles within active nuclei and extended lobes \cite{Rieger2008, Rieger2012}. And third, increasing the sample of $\gamma$-ray RGs will enable us to understand the contribution of nearby non-blazar AGN to the extragalactic $\gamma$-ray background.

PKS\,0625-354 (RA = 06$^\mathrm{h}$ 27$^\mathrm{m}$ 06.7$^\mathrm{s}$ DEC = -35$^\mathrm{d}$ 29$^\mathrm{m}$ 15$^\mathrm{s}$, J2000) is a radiogalaxy of type FRI located in the Abell Cluster 3392, with a redshift value z = 0.055 \cite{Jones2009}. It was detected by the {\it Fermi}-LAT in the high-energy (HE; 100 MeV $<$ E$ <$ 100 GeV) regime as a rather hard spectrum source $\Gamma_{\mathrm{3FGL}}$ = 1.88 $\pm$ 0.06 with a flux of F$_{1\mathrm{GeV}-100\mathrm{GeV}}$ = (1.43 $\pm$ 0.11) $\times$ 10$^{-9}$ ph cm$^{-2}$ s$^{-1}$ \cite{FermiCat}.

\section{H.E.S.S. data analysis and results}
H.E.S.S. (High Energy Stereoscopic System) is an array of five imaging atmospheric Cherenkov telescopes (IACT) located in the Khomas Highland in Namibia, which observes the gamma-ray sky with energies from tens of GeV up to around 100 TeV \cite{hess2}. 

\begin{figure}
\centering
\includegraphics[width=.4\textwidth]{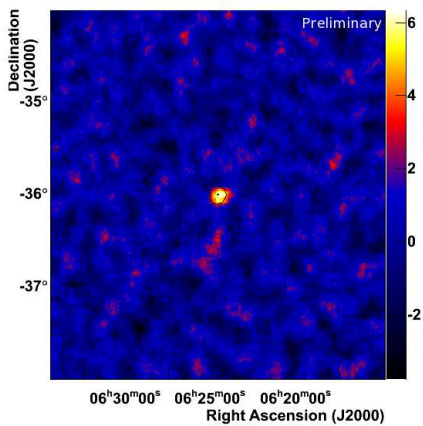}
\caption{Significance map centred on the position of PKS\,0625-354.}
\label{fig1}
\end{figure}

\begin{figure}
\centering
\includegraphics[width=.6\textwidth]{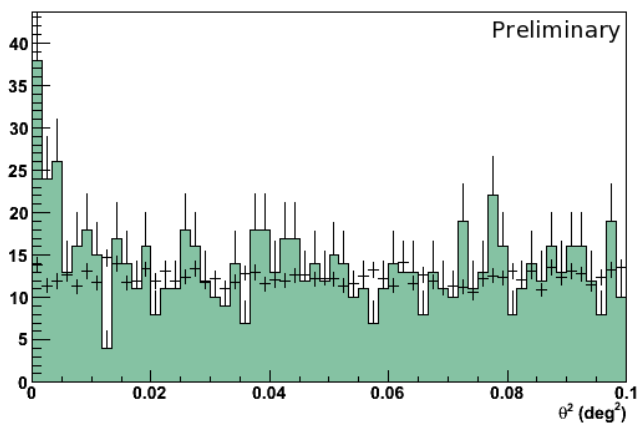}
\caption{Distribution of $\theta^2$ for on-source and normalized background. The excess is clearly visible in the region $\theta^2$ < 0.01 deg$^2$, corresponding to a statistical significance 6.1 $\sigma$.}
\label{fig2}
\end{figure}

PKS\,0625-354 was observed in 2012 with H.E.S.S. phase I array of four IACTs \cite{CrabPaper}. All data collected during these observations were taken in {\it wobble mode} with an offset of 0.5$^o$ from the source. In this analysis, 5.5 hrs of good quality, life-time corrected exposure with at least 3 telescopes have been used. These data were proceeded using the Model analysis chain with standard cuts applied \cite{model}. The total observed excess is 61 gamma-ray events, corresponding to a statistical significance of 6.1 standard deviations ($\sigma$). The significance map of PKS\,0625-354 is shown in Figure \ref{fig1}. Figure \ref{fig2} shows the on-source and normalized off-regions distributions as a function of squared angular distance from the source. The background is approximately flat and there is a point-like excess at small values of $\theta^2$, corresponding to the signal observed from PKS\,0625-354.

\begin{figure}
\centering
\includegraphics[width=.6\textwidth]{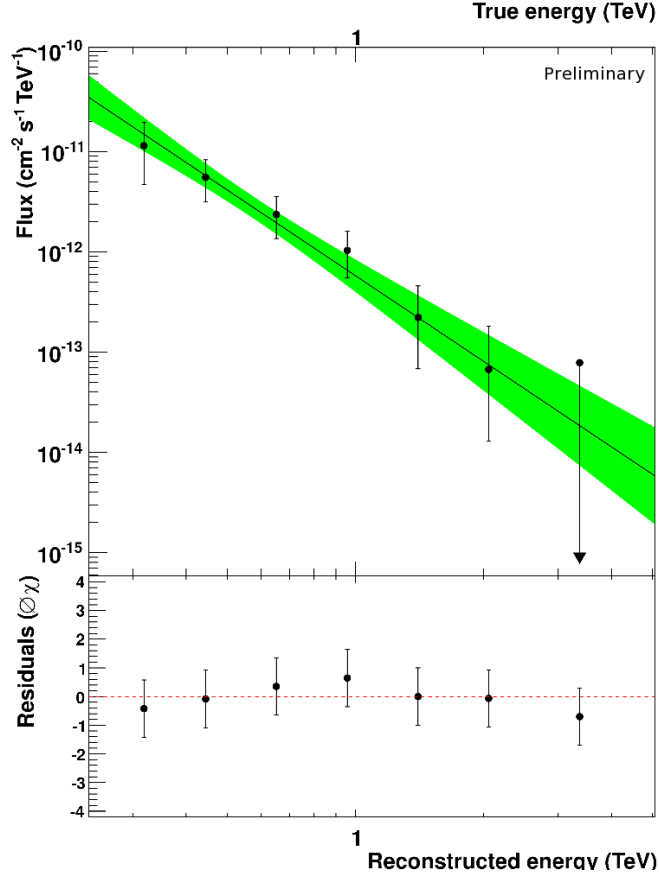}
\caption{VHE spectrum of PKS\,0625-354. The green region represents the 1-$\sigma$ confidence bounds of the fitted spectrum for power-law hypothesis.}
\label{fig3}
\end{figure}

The photon spectrum for the VHE data is shown in Figure \ref{fig3}. These data are well fitted ($\chi^2$/dof = 24.2/21) by a power-law function dN/dE = $N_0$(E/E${_0}$)$^{-\Gamma}$, with photon index $\Gamma$ = 2.8 $\pm$ 0.5, normalization constant $N_0$($E_0$) = (2.77 $\pm$ 0.70) 10$^{-12}$ cm$^{-2}$ s$^{-1}$ TeV$^{-1}$ (statistical errors only)  and decorrelation energy $E_0$ = 0.58 TeV, and it is the energy for which the correlation between the normalization constant and photon index is equal zero and hence the uncertainties are minimized. An integral flux above the decorrelation energy $I$ (> 580 GeV) = (8.7 $\pm$ 3.2)$\times$10$^{-13}$ cm$^{-2}$ s$^{-1}$, which correspond to $\sim$ 4\% of the H.E.S.S. Crab Nebula flux \cite{CrabPaper}, above the same threshold. As a cross-check, the observational data were analysed using independent analysis chain based on the event reconstruction algorithm - {\it ImPACT} \cite{Parsons}.

\section{Multiwavelength data analysis and results}
\subsection{High-energy observations with {\it Fermi}-LAT}
The Large Area Telescope LAT on board {\it Fermi} satellite is a pair-conversion gamma-ray detector, which operates in the high-energy regime (HE; 100 MeV $<$ E$ <$ 100 GeV) \cite{Atwood}. {\it Fermi}-LAT analysis was performed using data from 54622 to 56778 MJD (68 months), what corresponds to period between the start of the mission and end of April 2014. For the analysis the photons with a zenith angle $<105^\circ$ were selected. The binned maximum-likelihood method was applied. The Galactic diffuse background was modelled using the \verb|gll_iem_v05| map cube, and the extragalactic diffuse and residual instrument backgrounds were modelled jointly using the \verb|isotropic_iem_v02| template. All the sources from the {\it Fermi}-LAT Third Source Catalog inside the region of interest of PKS 0625-354 were modelled.
Two spectral models were tested with these data: power-law and a log-parabola. The likelihood test showed that the log-parabola model is preferred with the log-likelihood value of -253486 versus -253488 in case of power-law model and thus, the significance of the log-parabola with respect to  the power-law hypothesis is $\sim$ 2 $\sigma$. The fit parameters for a log-parabola model dN/dE = $N_0(E/E_b)^{-(\alpha+\beta \log{E/E_b})}$ are: prefactor - $N_0$ = (0.035 $\pm$ 0.012) 10$^{-9}$ cm$^{-2}$ s$^{-1}$ MeV$^{-1}$, index - $\alpha$ = 1.29 $\pm$ 0.18, curvature parameter - $\beta$ = 0.07 $\pm$ 0.02 and enery scale - $E_{B}$ = 100 MeV. 

\subsection{X-ray and UV observations with {\it Swift}}
The {\it Swift} Gamma-Ray Burst Mission \cite{Gehrels04}, launched in November 2004, is a multi-wavelength space observatory, equipped with three instruments: Burst Alert Telescope (BAT), X-ray Telescope (XRT)  and Ultraviolet/Optical Telescope (UVOT). PKS\,0625-354 was monitored with Swift in 4 observations with ObsIDs: 00039136001, 00049667002, 00039136002, 00049667001. 

The data were analysed using HEASoft package v.\,6.16 software\footnote{http://heasarc.gsfc.nasa.gov/docs/software/lheasoft} with CALDB v.\,20140120. 
In the case of X-ray observations all the events were cleaned and calibrated  using \verb|xrtpipeline| task. Data in the energy range of 0.3-10\,keV with grades 0-12 were analysed.  For the spectral studies, data were grouped using \verb|grappha| tool to have minimum 20 counts per bin  and the  spectra were fitted using XSPEC v.\,12.8.2.
Data were fitted with a single power-law model with Galactic hydrogen absorption value of $n_H=6.5\cdot 10^{20}$\,cm$^{-2}$ \cite{Kalberla05} fixed as frozen parameter.

Simultaneously with XRT observations  the source was monitored with UVOT. For each of ObsID, the instrumental magnitudes as well as  the corresponding fluxes are calculated following \verb|uvotsource| command,  taking into account all photons from a circular region within radius of 5''. 

\begin{figure}
\centering
\includegraphics[width=1.0\textwidth]{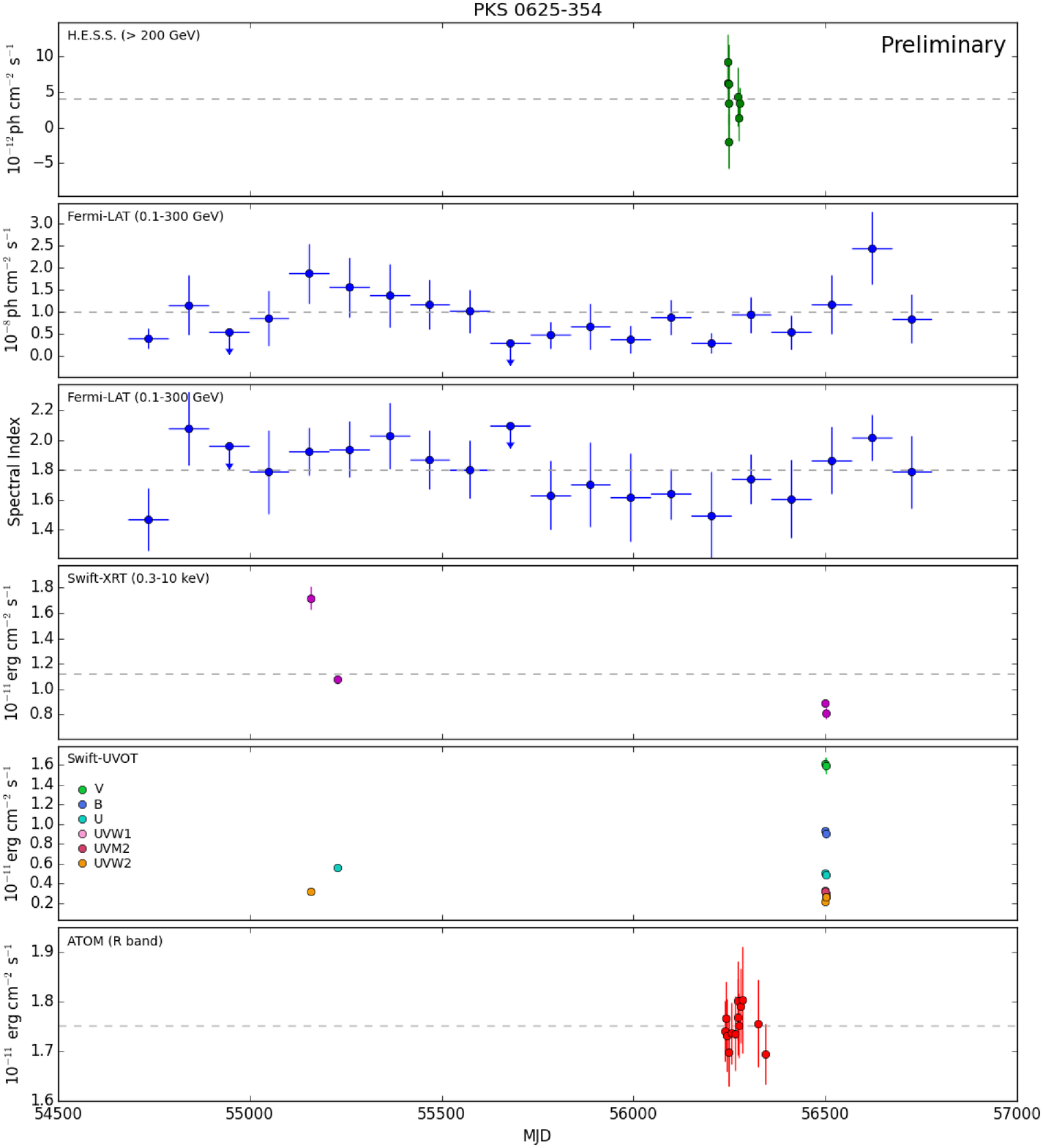}
\caption{Light curves of PKS\,0625-354 in different wavelengths. Top panel shows the H.E.S.S. lightcurve with nightly binning. The next two panels shows the HE flux changes and the {\it Fermi}-LAT inferred photon index changes. Another two panels are the X-ray and UV fluxes. Bottom panel are the ATOM observations in R-band.}
\label{fig4}
\end{figure}

\subsection{Optical monitoring with ATOM}
Extragalactic H.E.S.S. targets are monitored with the 75\,cm Automatic Telescope for Optical Monitoring (ATOM) located in Namibia at the H.E.S.S. site. The telescope is operational in Namibia since 2006 and monitored sources in UBVRI filters.  The detailed description of the instrument can be found in \cite{Hauser}. PKS\,0625-354 was monitored with ATOM in the R band during the period of  November-December 2012 and in February 2013. 

\section{Conclusions}
Observations carried with the H.E.S.S. phase I array in 2012 have established PKS\,0625-354 as VHE gamma-ray source, which enlarges the extragalactic non-blazar source class. The available multiwavelength data as depicted in Figure \ref{fig4}, although not simultaneous, will be used in future in a more detailed SED study and offers the possibility to discriminate between different SED models. No significant variability of the VHE flux was noticed during the HESS observations period (see Fig. \ref{fig4}). In the HE range, analysis of observations simultaneous to H.E.S.S. ones yields only flux upper limit, although if the total {\it Fermi}-LAT data set is considered there is a hint of variability, as seen in Figure \ref{fig4}. A constant-flux yields the fit of the probability of P($\chi^2$/dof = 30.128/17) = 0.81. The source is also variable in X-ray range ({\it Swift}/XRT) although there is no-simultaneous monitoring with VHE observations. In the optical band ATOM observations show almost constant flux in R band. Only 3-4 pointing observations in optical/UV energy bands do not allow either to claim or to exclude variability in these regimes.

\acknowledgments{The support of the Namibian authorities and of the University of Namibia in facilitating the construction and operation of H.E.S.S. is gratefully acknowledged, as is the support by the German Ministry for Education and Research (BMBF), the Max Planck Society, the German Research Foundation (DFG), the French Ministry for Research, the CNRS-IN2P3, and the Astroparticle Interdisciplinary Programme of the CNRS, the U.K. Science and Technology Facilities Council (STFC), the IPNP of the Charles University, the Czech Science Foundation, the Polish Ministry of Science and Higher Education, the South African Department of Science and Technology and National Research Foundation, and by the University of Namibia. We appreciate the excellent work of the  technical support staff in Berlin, Durham, Hamburg, Heidelberg, Palaiseau, Paris, Saclay, and in Namibia in the construction and operation of the equipment.}

\end{document}